# City Brain, a New Architecture of Smart City Based on the Internet Brain

1st *Liu Feng*

Research Center on Fictitious Economy & Data Science,

CAS,

Beijing, China

*Abstract*—In the ten years after the Smart City was put forward, there are still problems like unclear concept, lack of top-down design and information island. With the further development of the Internet, the brain-like architecture of the Internet is becoming clearer and clearer. As a product of combination of city buildings and the Internet, the Smart City will also have a new architecture, and the City Brain thus appears. Based on the Internet Brain, this paper describes how to construct the Smart City in the form of brain-like tissue, and how to evaluate the construction level of the Smart City (City IQ) relying on the Big SNS (city neural networks) and city cloud reflex arcs.

I. Put-forward of the Smart City and Existing Problems

On November 6, 2008, Peng Mingsheng, the president and CEO of IBM U.S. delivered a speech titled *A Smarter Planet: the Next Leadership Agenda* in the conference of the Council on Foreign Relations in New York City [1] Thus, the concept of Smart Planet was definitely put forward for the first time, the goal of which is to enable the society to advance more smartly, the human to survive more smartly and the Planet to run more smartly.

On this basis, the concept of Smart City came into being. And in the following eight years, it was taken as the goal of the world's city construction and has been vigorously promoted[2]. However, there are still problems like unclear concept, blind speculation, lack of top-down design and fragile foundation. Relevant experts have pointed out that some of the Smart City construction projects are lack of top-down design and overall co-ordination, with a big difficulty in coordination and docking, especially in some vertical sectors, where the information system is only used within internal departments, lacking of data sharing and application between different departments. Besides, the evaluation standards are not unified, which shows a big gap relative to the citizens' demands [3].

II. Generation of the Internet Brain

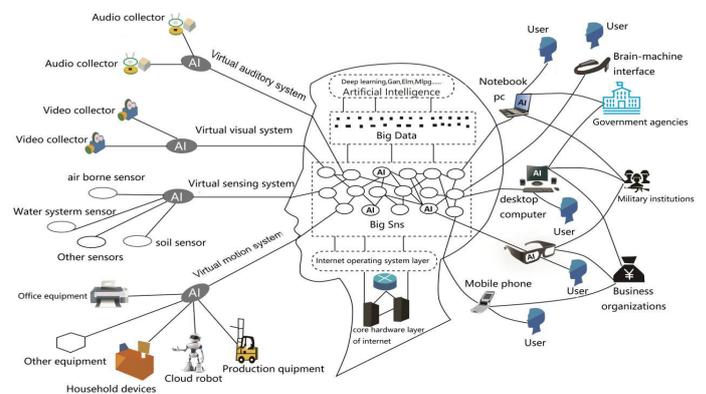

Fig. 1 Architecture Chart of the Internet Brain

The influence of the Internet Revolution[4] on the human has gone far beyond the Industrial Revolution. Unlike the Industrial Revolution which strengthened the power of and expanded the vision of the human[5], the Internet greatly enhanced the human wisdom and enriched our knowledge while the wisdom and knowledge are precisely the most closely related to the brain. The technological breakthrough of the Internet makes mankind on the eve of the new scientific revolution.

Since the birth of the Internet in 1969, the human have been performing innovation in the field of the Internet from different aspects[6], but there is no unified plan about the structure of the Internet to be constructed. With the rapid development of the cutting-edge technologies like artificial intelligence (AI), Internet of things, big data, cloud computing, robot, virtual reality, industry Internet and so on, their cooperative effects enable the brain-like architecture of the Internet becomes clearer and clearer.

In 2008, the research team formed with the author of this paper as well as Peng Geng and Liu Ying from the Chinese Academy of Sciences published a paper, which put forward that "the Internet will evolve toward the direction highly similar to the human brain, and it will have its own memory nervous system, central nervous system and autonomic nervous system in addition to the visual, auditory, tactile sense and motor nervous systems. On the other hand, the human brain had had all the Internet functions through evolution for at least tens of thousands of years, and the continuous development of the Internet will help neurologists to reveal the secret of human brain. Scientific experiments will prove that the brain also has a search engine like Google, a SNS system like Facebook, an address encoding system like IPv4, and a routing system like Cisco...". This brain-like architecture of the Internet has been named the Internet Brain(Fig. 1)[7].

III. City Brain, a New Architecture of Smart City based on the Internet Brain

A Definition of City Brain

It should be said that the Smart City is the result of the spread and deepening of the Internet in cities after it has developed to a certain degree. Therefore, the development and evolution laws of the Internet should not be ignored in the construction of the Smart City. As a product of combination of the Internet Brain and the construction of cities, the Smart City will inherit the basic characteristics of the Internet Brain, so it is defined as follows:

The City Brain is a new architecture of the Smart City based on the model of the Internet Brain. Under the support of the city central nervous system (cloud computing), the city sensory nervous system (Internet of Things), the city motor nervous system (Industry 4.0, Industrial Internet) and the city nerve endings (Edge Computing), a city can achieve the human-human, human-things and things-things information interaction through the city neural network (Big SNS) and achieve the rapid smart response to city services through the city cloud reflex arcs, so as to promote the organic integration of all components of a city, realizing the continuous progress of city wisdom. Such a brain-like smart city architecture is called "City Brain".

B How to Construct the City Brain

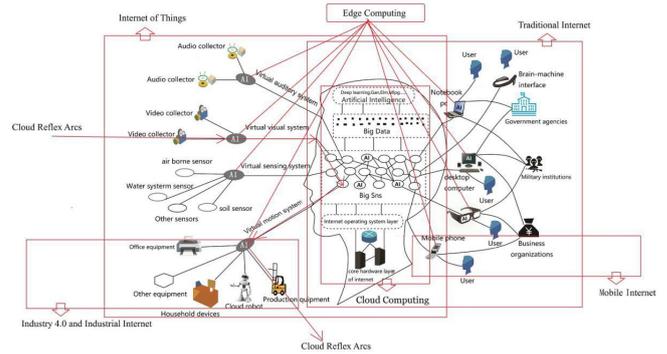

Fig. 2 Architecture Chart of the Construction of the City Brain

In the construction of the Smart City, the key technologies include Internet of things, Big Data, Cloud Computing, AI, Industry 4.0, Industrial Internet and robots. At the same time, the new technology frameworks like BIG SNS and cloud reflection arc are also emerging. And what's the relationship between these cutting-edge technologies and concepts and the City Brain, and how to realize them in the construction of the City Brain? A detailed description is given below (Fig. 2).

1. The construction of Internet of Things for the City Brain involves the construction of a city's somatosensory nervous system and motor nervous system, mainly consisting of the sensors and smart driving systems, cloud robots, unmanned aerial vehicle, 3D printing systems and smart manufacturing systems distributed in the fields like enterprises, construction, finance, transportation and energy of the city[8].

2. The construction of cloud computing for the City Brain refers to the construction of a city's central nervous system. The central nervous system controls all the other nervous systems of the city through the server, network operating system, neural network (Big SNS), Big Data and artificial intelligence algorithm based on Big Data. It should be noted that: since the City Brain is a subset of the Internet Brain, the construction of the Smart City is not isolated, and the architectures of different City Brains are often crossed. For example, a city's cloud computing hardware facility is likely located in another city, and a city's Big Data may be distributed and stored in different cities[9].

3. The construction of Big Data for the City Brain is essentially the valuable information that all the nervous systems of the City Brain transmit and accumulate in the operation process . Such information is respectively from the data generated from the residents' life, the enterprises' operation and the government's management, or from the city's buildings and traffic vehicles, or even from the sensors distributed in the soil, air and waters of the city, which is the basis for the City Brain to actually generate wisdom[10].

4. The Industry 4.0[11] and the Industrial Internet[12] of the City Brain is a product from the development of the city's motor nervous system, and it will be a huge component of the City Brain in the future, including smart driving systems, cloud robots, unmanned aerial vehicles, 3D printing systems, smart manufacturing systems and so on, which can help the city residents and managers to operate and construct the city through extension movement and mechanical operation .

5. The construction of edge computing[13] is the development and growth of the nerve endings of the City Brain. It integrates the AI technologies and the chips with AI technologies into the sensors, cameras, smart terminals, smart cars, smart manufacturing equipment, buildings, industrial robots and other equipment distributed across the city, the goal of which is to make the endings of the sensory nervous system and the motor nerve system of the City Brain smarter and more robust.

6. The construction of mobile Internet is to enrich the nerve fiber types of the City Brain so that city residents, enterprises and government agencies are linked into the City Brain more conveniently, with less geographical restrictions. This requires the continuous technological upgrading and infrastructure construction of the city's communications operators. The construction of nerve fibers for the City Brain is an important guarantee to ensure the operation of the Smart City, and the key to connect cities as a whole and ensure the timely and accurate operation of the city's cloud reflex arcs[14].

7. The construction of AI for the City Brain is the catalyst and soul to enhance the wisdom of the Smart City . The AI is not only combined with Big Data through algorithms (e.g., deep learning and machine learning), but also applied into the nerve endings, neural networks and smart terminals of the City Brain(e.g., AI sensors, AI mobile phones, AI intelligent production equipment, AI user assistant, etc.). The combination of AI and the Smart City will enable the capabilities of all nervous systems of the Internet Brain to improve synchronously.

8. The construction of Big SNS is the development of the neural network of the City Brain. It should be said that the neural network of the City Brain is one of the two most important factors in the construction of the Smart City . The social network has always been considered as a human-human interaction community on the Internet. With the emergence of new phenomena like Internet of Things, cloud computing and Big Data, however, the form of the social network will certainly change as well. When the Internet of things, Industry 4.0, Industrial Internet and social networks are integrated, every building, every car, every scenic spot, every mall and every electrical appliance will respectively be able to have an account on the SNS website, through which their real-time information will be issued automatically, and they can also interact with other "persons" and "things" . Then, the definition of social networks will no longer just be the human-human interaction, but bigger social networks involving human-human, human-things, and things-things interactions, which we call  the neural network  or "Big Social Networks" (Big SNS) (see Fig. 3).

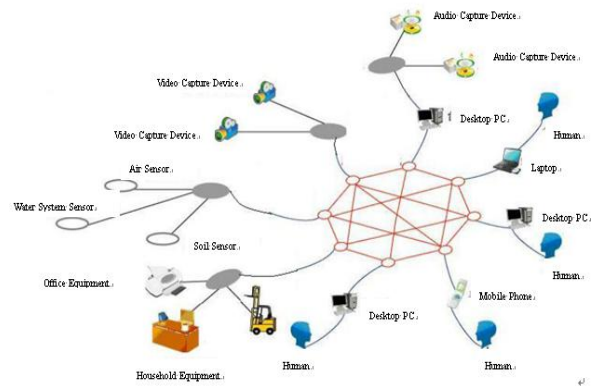

Fig. 3 Architecture Chart of Big SNS

The close combination of this Big SNS and the Smart City is the key to successful construction of the City Brain. No matter residents, enterprises, government agencies, street lamps, vehicles or plants should be added into the Big SNS in the form of city neurons. The interaction, the combination and the interlinkage of such city neurons will make the City Brain smarter.

9. The construction of the cloud reflex arcs for the City Brain is the second most important factor in the construction of the Smart City. As we know, the nerve reflex is one of the most important neural activities of the human nervous system [15] (see Fig. 4). It include the sensor, afferent nerve fibers, nerve center, efferent nerve fibers and effector.

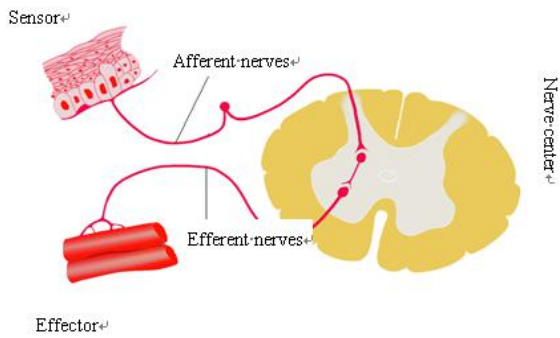

Fig. 4 Diagram of Neural Reflex Arc[16]

The reflex arc is the structural basis of reflex activity and is the entire path of excitement circulating in the nervous system from the moment the stimuli are received by the body to the time when the reaction occurs. The reflex is generally achieved through a complete reflex arc. A complete reflex arc consists of five basic components, i.e. receptor, afferent nerve, nerve center, efferent nerve and effector. Knee-jerk reflex is the easiest reflex activity that we are familiar with. Then, is there similar mechanism in Internet cloud brain-based artificial intelligence applications?

Corresponding to the neural reflex arc of human body, the Internet cloud reflex arcs mainly consists of the following three aspects: Firstly, the receptor of the cloud reflex arcs is mainly composed of networked sensors (including cameras); secondly, the effector of the cloud reflex arcs are mainly composed of networked office equipment, smart manufacturing, smart driving, smart medical and so on; and, thirdly, the central nerve of the cloud reflex arcs is the central nervous system (Cloud Computing+ Big Data + Artificial Intelligence) of the Internet Cloud Brain. Edge computing will enhance the intelligence degree and response speed of the receptors and effectors of the cloud reflex arcs.

The cloud neural reflex arc is the basis of the reflex phenomenon in the Internet Cloud Brain, and it has appeared widely around us today. The Internet neural reflexes initiated from all over the world are continually occurring and disappearing at every moment. For example, the car sensor has found a thief, sends a message to the car owner, and the owner will rush to catch the thief; the humidity sensor has sensed the increased air humidity and there are signs of rain, it will inform the field excavation equipment to turn on the rain-proof devices.

Definition of the cloud reflex arc of the city brain:

In the brain-like intelligent architecture of a smart city, the cloud reflex arc is namely a complete intelligent reaction chain required for the operation of a city, which is realized by cloud computing, with the big data and artificial intelligence as the central nervous system, various types of communication lines as afferent nerve fibers and efferent nerve fibers, the networked electronic sensors like visual sensors, auditory sensors, and touch sensors as well as human perception participation as the receptor, smart devices like networked robots, smart office, smart furniture, smart manufacturing, smart driving and smart medical devices as well as human operation participation as the effector.

There have already been many cases of neural reflex arcs in the Smart City. From the current view, the City Brain involves a total of nine types (see Fig. 5)[17].

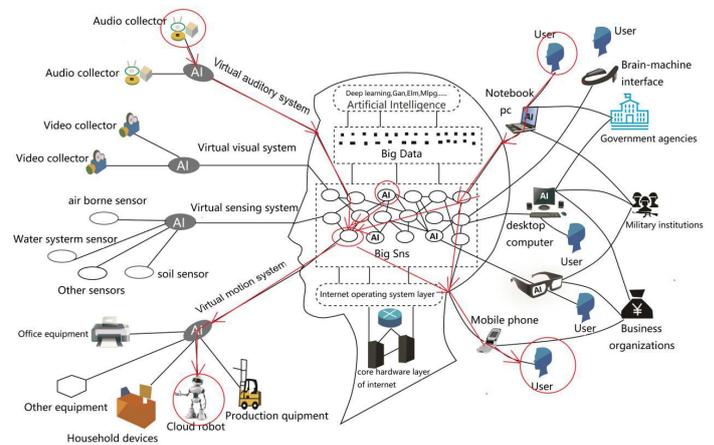

Fig. 5 Cloud Reflex Arc

The first type is the cloud reflex arcs from sensors to smart devices (A-> F in the figure): For example, in a building, when the temperature sensor detects a room temperature rise exceeding a certain temperature and the gas sensor detects an increase of $CO_2$ concentration in the room, an alarm message will be sent to the server center through the Internet line. The server will issue a command to the fire extinguishing robot of the building, and then the fire extinguishing robot will operate the water gun to put out the fire.

The second type is the cloud reflex arcs from sensors to humans (A-> B in the figure): For example, in a building, the temperature sensor detects a room temperature rise of more than 100 degrees and the gas sensor detects an increase of $CO_2$ concentration in the room, an alarm message will be sent to the server center through the Internet line. The server will send the message to the fire brigade nearby, and the fire brigade will dispatch firefighters to the building to put out the fire.

The third type is the cloud reflex arcs from sensors to smart programs (A-> D in the figure): For example, in a building, the temperature sensor detects a room temperature rise of more than 100 degrees and the gas sensor detects an increase of $CO_2$ concentration in the room, an alarm message will be sent to the server center through the Internet line. The server will send the message to the AI neurons in the Internet neural network, i.e. the smart program in the Big SNS, and then the smart program will decide the danger level and whether to report the case.

The fourth type is the cloud reflex arcs from smart programs to smart devices (D-> F in the cloud reflex architecture diagram): For example, an automated monitoring program running on the Internet server detects the capacity changes of the server data space of the cloud computing room located in city or suburb. When the program finds that the data space is full, it will send an alarm message to the Internet center server, and then, the center server will issue a command to start the spare device in the cloud computer room to expand the data space.

The fifth type is the cloud reflex arcs from smart programs to humans (C-> E in the figure): For example, an automated monitoring program running on the Internet server detects the capacity changes of the server data space of the cloud computing room located in the suburb. When the program finds that the data space is full, it will send an alarm message to the Internet center server, and then, the center server will send a message or an e-mail to remind the personnel on duty in the cloud computing room to start the spare device to expand the data space.

The sixth type is the cloud reflex arcs from smart programs to smart programs (C-> D in the figure): This type of nerve reflex arcs can be regarded as dialogues between cloud artificial intelligence systems. For example, an automated monitoring program running on the Internet server detects the capacity changes of the server data space of the cloud computing room located in the suburb. When the program finds that the data space is full, it will send an alarm message to the Internet center server, and then, the center server will issue a command to the maintenance program of the cloud computing room to stop writing data to the data space, to avoid overload of the data space.

The seventh type is the cloud reflex arcs from humans to smart devices (B-> F in the figure): For example, in a building, the watchman on duty in the monitoring computer room discovers flames and smoke in an office, he/she can press an alarm button to send an alarm message to the server center through the Internet line, and the server will issue a command to the fire extinguishing robot of the building. And then, the fire extinguishing robot will operate the water gun to put out the fire.

The eighth type is the cloud reflex arcs from humans to humans (B -> E in the cloud reflex architecture diagram): For example, in a building, the watchman on duty in the monitoring computer room discovers flames and smoke in an office, he can press an alarm button to send an alarm message to the server center through the Internet line, and the server will send a message to the fire brigade nearby, and the fire brigade will dispatch firefighters to the building to put out the fire.

The ninth type is the cloud reflex arcs from humans to smart programs (B-> D in the cloud reflex architecture diagram): For example, in a building, the watchman on duty in the monitoring computer room discovers flames and smoke in an office, he can press an alarm button to send an alarm message to the server center through the Internet line. And then, the server will send the information to the AI neurons in the Internet neural network, i.e. the smart program in the Big SNS and then the smart program will decide the danger level and whether to report the case.

## IV. Methods to Evaluate the Development Levels of City IQ and City Brain

A Definition of (Smart) City IQ

In the previous studies, it was put forward that the essence of the Smart City construction is the combination of the Internet Brain architecture and the city construction. We know that there is IQ evaluation for the brain[18]; for the brain-like Smart City architecture, therefore, the City IQ can also be used to express the development level of a city's wisdom .

It should be noted that the core of the Smart City is the city neural network and the city cloud reflex arcs. The cloud computing, Internet of things, Industry 4.0 and Big Data are all used to support them. However, we should also note that a city's construction is not isolated, and the cloud brain architectures of different smart cities are often crossed. For example, a city's cloud computing hardware facilities are likely located in another city and the Big Data of a city may be distributed and stored in different cities.

The regional scale and populations of different cities are quite different. Therfore, we should not investigate the indexes affecting the Smart City IQ only depending on the city's development levels in cloud computing, big data and Internet of Things, but focus on the city neural network coverage and the construction of city cloud reflex arcs which are not related to the scale and population of the city.

Based on above research, the City IQ presented in this paper is defined as an Internet Brain model based on the Smart City, performing the comprehensive evaluation on the two core elements, i.e. the neural networks (City Big SNS) and the cloud reflex arcs of the target city under the support of the city central nervous system (cloud computing), the city sensory nervous system (Internet of Things), the city motor nervous system (Industry 4.0, Industrial Internet) and the nerve endings (Edge Computing), in order to measure the intelligence development of the evaluated city at the time of measurement. The result is namely the City IQ of the city at the measurement time.

B Evaluation Method of City Neural Networks (City Big SNS)

In the city neural network (City Big SNS), a total of four level-2 indexes are set up in the test scale, including :

1. The stability (robustness) of city neural networks, which represents the stability of the hardware infrastructures and software systems of a city's neural network, and can be measured by the annual failure rate of the system.

2. The uniformity of city neural networks. Currently, the Smart City contains so many types of systems which cannot be connected with each other, reducing the uniformity of the architectures of city neural networks. Therefore, the uniformity of a city's neural networks is evaluated through the number of smart city systems and the establishment of the Big SNS in a city.

3. The coverage of city neural networks. This index is mainly used to evaluate the percentages of population, business organizations, government agencies and city equipment of the city that have been unified into a Big SNS and achieved information exchange.

4. The activeness of city neural networks. This index is mainly used to evaluate the activeness of information transmission and exchange of the population, business organizations, government agencies and city equipment which are linked to a city Big SNS.

C Evaluation Method of City Cloud Reflex Arcs

The construction of city cloud reflex arcs reflects the types of wisdom-related services that may be provided by a city and the city's response speed in the process of providing such services. The more the types of city reflex arcs are, the faster the response speed will be, and the higher the smart level will be as well. For the city cloud reflex arcs, N level-2 indexes and two level-3 indexes (i.e. robustness and response speed) are set up in the test scale.

As the construction of the Smart City involves various types of cloud reflex arcs, the contents of level-2 indexes include security cloud reflex arcs, finance cloud reflex arcs, traffic cloud reflex arcs, energy cloud reflex arcs, education cloud reflex arcs, medical service cloud reflex arcs, tourism cloud reflex arcs, retail cloud reflex arcs and so on .

The types of these cloud reflex arcs change with the development of the Smart City. In convenient for standardization and measurement, a standard library of the cloud reflex arcs types of the city may be developed, where the types of cloud reflex arcs may be added or deleted on a yearly basis. Based on the above discussion, the first edition of City IQ Test Scale 2017 forms as below (Table 1):

D Evaluation Method of City Cloud Reflex Arcs

According to the measurement method presented in this paper, we can measure a city's IQ mainly from the perspective of the construction of city neural networks (city Big SNS) and city cloud reflex arcs (see Table 1), and develop an annual (Smart) City IQ Ranking and Research Report in the future.

Table 1 City IQ Test Scale

| City IQ Test Scale (Version 2017) | | |
|---|---|---|
| Level-1 Indexes | Level-2 Indexes | Level-3 Indexes |
| City neural networks (City Big SNS) | Completeness of city neural networks | |
| | Uniformity of city neural networks | |
| | Coverage of city neural networks | |
| | Activeness of city neural networks | |
| City cloud | Security cloud reflex arcs | Response |

| reflex arcs | | speed of reflex arcs |
| --- | --- | --- |
| | | Stability (robustness) |
| | Finance cloud reflex arcs | Response speed of reflex arcs |
| | | Stability (robustness) |
| | Traffic cloud reflex arcs | Response speed of reflex arcs |
| | | Stability (robustness) |
| | Logistics cloud reflex arcs | Response speed of reflex arcs |
| | | Stability (robustness) |
| | Energy cloud reflex arcs | Response speed of reflex arcs |
| | | Stability (robustness) |
| | Education cloud reflex arcs | Response speed of reflex arcs |
| | | Stability (robustness) |
| | Community cloud reflex arcs | Response speed of reflex arcs |
| | | Stability (robustness) |
| | Medical service cloud reflex arcs | Response speed of reflex arcs |
| | | Stability (robustness) |
| | Tourism cloud reflex arcs | Response speed of reflex arcs |
| | | Stability (robustness) |
| | Retail cloud reflex arcs | Response speed of reflex arcs |
| | | Stability (robustness) |
| | Agricultural trade cloud reflex arcs | Response speed of reflex arcs |
| | | Stability (robustness) |
| | Environmental protection cloud reflex arcs | Response speed of reflex arcs |
| | | Stability (robustness) |
| | To be further added according to the study results. | |

## V. Conclusion and perspectives

In this paper, we will discuss the current problems of the Smart City, including unclear concept, information island and excessively decentralized and separated construction. On the basis of the theoretical architecture of the Internet Brain, this paper puts forward the brain-like model for the construction of the Smart City, namely the City Brain. Compared to the previous research of the Smart City which mainly focused on the fields of the Internet of Things, cloud computing, big data and artificial intelligence (AI), the City Brain focuses on the construction of big SNS and cloud reflex arcs. It proposes that only when the big SNS forms, can all the components of the Smart City be combined together organically. And only when the city cloud reflex arcs appear, can the intelligent services of the Smart City really function. The evaluation on these two core points has become the basis for judging the development level of the Smart City. After the City Brain architecture is proposed, there are two aspects of work to be completed. The first one is to study the implementation of the Big SNS construction, namely, how to build neuronal functions, permissions and interconnection relationship of Big SNS; and the second one is to study the composition of the city cloud reflex arcs, namely to determine what intelligent technologies and Internet functions are required to form necessary functional reflection arcs in the construction of a smart city, to ensure the efficient operation of the city; and what cutting-edge technologies are required to build the cloud reflection arcs based on the unique development fields of different cities, to help the city achieve smart city functions with their own characteristics.